\documentclass{kluwer}
\input{epsf.sty}
\relax
\def\eps@scaling{.95}
\def\epsscale#1{\gdef\eps@scaling{#1}}
 
\def\plotone#1{\centering \leavevmode
    \epsfxsize=\eps@scaling\columnwidth \epsfbox{#1}}
 
\def\plottwo#1#2{\centering \leavevmode
    \epsfxsize=.45\columnwidth \epsfbox{#1} \hfil
    \epsfxsize=.45\columnwidth \epsfbox{#2}}

\begin{document}
\begin{article}
\begin{opening}
\title{The Resolved Red Giant Branches of E/S0 Galaxies}            

\author{Regina E. \surname{Schulte-Ladbeck}\email{rsl@phyast.pitt.edu}}
\author{Igor O.  \surname{Drozdovsky}\thanks{Astronomical Institute, St. Petersburg University, Russia}} 
\author{Mich\`ele \surname{Belfort}}
\institute{Department of Physics \& Astronomy, University of Pittsburgh, USA}
\author{Ulrich \surname{Hopp}} 
\institute{Universit\"ats-Sternwarte M\"unchen, F.R.G.}


\begin{abstract} 

Formation paradigms for massive galaxies have long centered
around two antipodal hypotheses --- the monolithic-collapse and the
accretion/merger scenarios. Empirical data on the stellar contents of galaxy halos is crucial
in order to develop galaxy formation and assembly scenarios which
have their root in observations, rather than in numerical simulations. Hubble Space
Telescope (HST) has enabled us to study directly individual stars in the nearby E/S0 galaxies
Cen~A, NGC~3115, NGC~5102, and NGC~404. We here present and discuss 
HST single-star photometry in V and I bands. Using color-magnitude
diagrams and stellar luminosity functions, we gauge the galaxies' 
stellar contents. This can be done at more than one position
in the halo, but data with deeper limiting magnitudes are desired
to quantify the variation of metallicity with galactocentric radius. 
We here compare the color distributions of red giant stars with stellar isochrones, and we
intercompare the galaxies' halo populations, noting that their total absolute 
V magnitudes cover the range from about -21.5 to -17.5. 
In the future, we plan to model the stellar metallicity distributions with the aim to constrain
chemical enrichment scenarios, a step towards unravelling the evolutionary 
history of elliptical and lenticular galaxies.

\end{abstract}


\end{opening}

\section{Introduction}
The stellar halos of galaxies contain the fossil
record of their formation and assembly. The origin of the Milky Way's
stellar halo has long been a topic of intense investigation.
Tests of formation scenarios for the halo have centered on two extreme
viewpoints --- the monolithic collapse model of Eggen, Lynden-Bell \&
Sandage (1962), and the accretion model of Searle \& Zinn
(1978). The Searle \& Zinn picture has gained momentum
in recent years because of observational evidence for halo assembly
from small building blocks (e.g., the tidal disruption by the Milky Way
of the Sagittarius dwarf galaxy, Ibata, Gilmore \& Irwin 1995).
On the theoretical side, the Searle \& Zinn idea has also become a
favorite ``Denkmodell" because it bears a strong resemblance to the popular hierarchical
galaxy formation scenario characteristic of cold dark matter (CDM)
cosmological models (e.g., White \& Rees 1978).  

It is our goal to determine stellar metallicities for halo field stars, for at least
two radial positions within each galaxy, and to compare the metallicity distribution 
functions (MDF) derived from observations, with predictions from chemical enrichment models.

\section{Observations \& Results}

To this end, we are in the process of reducing and analyzing HST WFPC2 observations
in the equivalent of the V (F606W) and the I (F814W) filters.
Note that the V-I color of the red giant branch (RGB)
has a much greater sensitivity to metallicity than it has to age, making 
it a good stellar metallicity indicator (although there is some degeneracy). 
We here suppose that the halo stars have ages of 10~Gyr, and compare with
Girardi et al. (2000) isochrones of different metallicity, readily available in the F606W, F814W plane. 
Spectroscopy of individual halo stars will be needed to break the age-metallicity degeneracy 
inherent to broad-band RGB colors.

\begin{figure}
\vspace{--0.5cm}
\plottwo{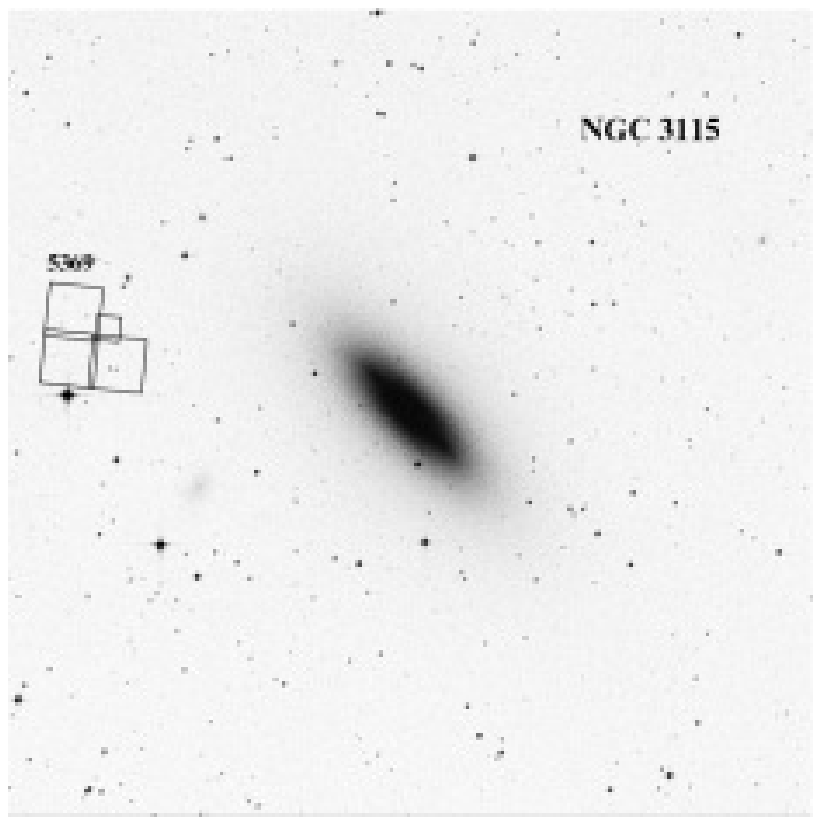} {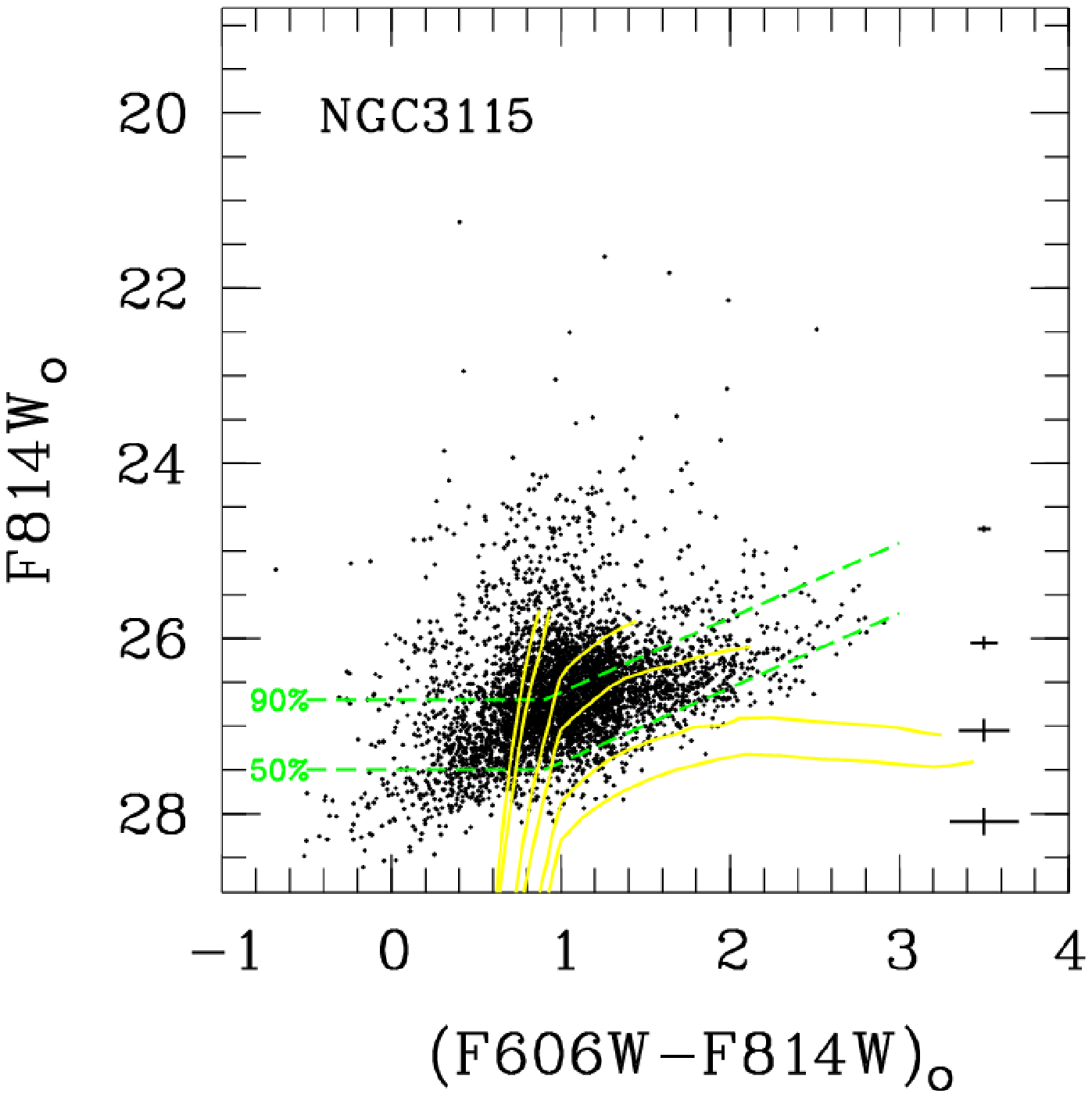}
\caption{HST observations of NGC~3115. The figure to the left illustrates the location
of the pointing of the WFPC2 overlayed on a DSS image of the galaxy. The figure to the
right shows the CMD in instrumental magnitudes in the Vegamag system. 
It contains 4,567 objects classified as point sources, but note some are foreground stars
in our Galaxy. Errorbars are given to the right of the data distribution. The dashed lines 
indicate the 90\% and 50\% completeness levels. Isochrones from Girardi et al. (2000) 
are overplotted as solid lines assuming a distance modulus of 29.72, an age of 10~Gyr, 
and for metallicities (from left to right) of 2\%, 5\%, 21\%, 42\%, 100\%
and 158\% of Solar. They show the data give access to the low-metallicity stars. 
The photometric limits cut off a potential high-metallicity stellar component.}
\end{figure} 

We restricted our work to observations in F606W and F814W. In particular,
we did not include data in the F555W filter, the more standard V filter,
or other filters which could be transformed to V and I, because we wish to assemble
a homogeneous database where direct intercomparison is possible in the observational plane.

At present, we are working on GO dataset 5905 for Cen~A (NGC~5128). 
The dataset 5369 is used for NGC~3115 (Fig.~1). NGC~5102 has two useful observations, in programs
8601 and 6252; and so has NGC~404, in programs 8106 and 5369.

\begin{figure}
\vspace{-1cm}
\hspace{-2cm}
\plotone{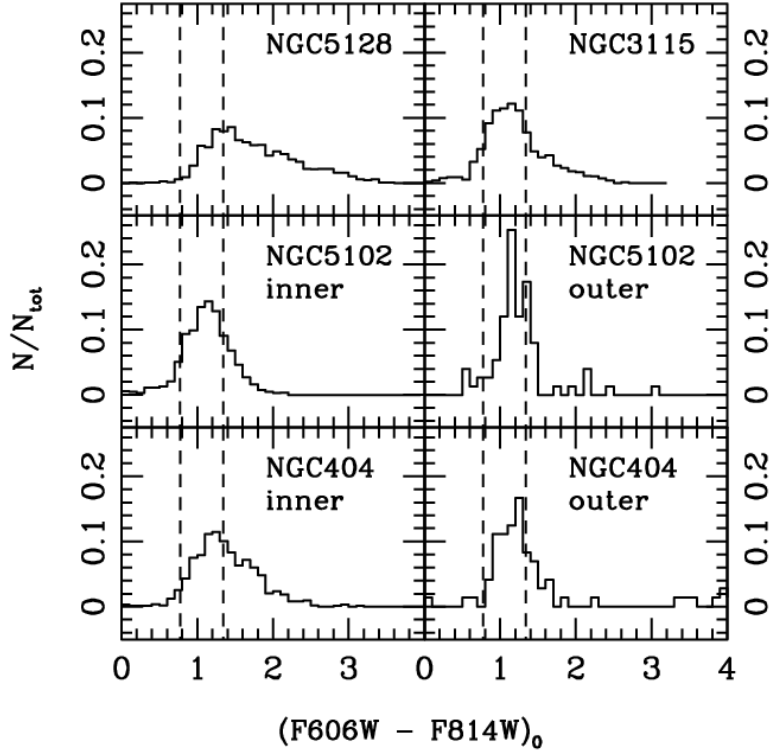} 
\caption{The normalized color distribution of stars from 0.3 to 0.7 magnitudes below the
tip of the RGB is shown for all fields. The dashed lines
correspond to the colors of the 1\% and 42\% of Solar metallicity isochrones. 
The photometric limits cut into a potential high-metallicity stellar component
redward of the 42\% Solar fiducial line for the NGC~3115 data, as well as for the inner halo
pointings of NGC~404 and NGC~5102. Nevertheless, a ``red tail" of stars is evident. The
red tail is missing in the deeper exposures of the outer halo fields of NGC~404 and NGC~5102; the metallicity
spread appears to be smaller here, but retains the peak value of the inner regions.}
\end{figure}

The data differ in exposure time and radial positioning. This affects the total number of resolved point sources. 
Stellar densities/crowding also vary with radius. We are carefully assessing the photometric errors and the completeness 
limits of the datasets. 

Fig.~2 shows preliminary results for color distributions at 0.5~mag. below the tip of the RGB. 
The data were not corrected for incompleteness. Completeness
is good to the 42\% Solar fiducial; it varies for redder/more metal rich stars from case to case.
We used the Kolmogorov-Smirnov statistic to investigate similarities and differences in the color
distributions with (F606W-F814W)$_0$$<$1.45. In general, the probability is high for the distributions
to be different. The distributions of NGC~404 inner \& outer, and NGC~404 outer \& NGC~5102 outer, on the
other hand, are similar to one another.

\section{Conclusions}

It remains difficult to obtain photometry of individual red giant stars in the halos of distant galaxies. 
However, the results presented here indicate the pursuit of deeper photometry is meritorious.
There are clear differences between the four galaxies. For example, we are sensitive to
the relative stellar contents at blue colors, where the data are highly complete for
all pointings.  The differences in color distributions could indicate a difference in the contents of 
low-metallicity stars. There is also a suggestion of a metallicity
gradient in the halo of NGC~5102, in the sense that the inner pointing contains
stars with a wider range of metallicity than the outer pointing. Additional data in the outer
halos of E/S0 galaxies, for which a large field of view is more important than extremely high spatial 
resolution, can be obtained with large ground-based telescopes, but are costly in terms of integration times
in V. 

\begin{acknowledgements}
We made extensive use of the HST archive and of the NED database. 
RSL thanks the Department of Physics \&
Astronomy for a leave of absence, and the MPE Garching for hosting her visit. ID acknowledges 
financial support from the University of Pittsburgh, Office of Research. 
UH acknowledges support from the SFB 375 of the DFG. 
\end{acknowledgements}

{}

\end{article}
\end{document}